\documentclass[prl,floatfix,twocolumn,showpacs]{revtex4} 
\usepackage{graphicx}
\def\bea{\begin{eqnarray}}
\def\eea{\end{eqnarray}}
\def\be{\begin{equation}}
\def\ee{\end{equation}}

\def\S{\mbox{\bf S}}

\def\et{{\it et al.}}

\def\kag{{\it kagom\'e }}
\begin{document}
\author{Andreas L\"auchli}
\author{Didier Poilblanc}
\affiliation{
  Laboratoire de Physique Th\'eorique, CNRS-UMR 5152, Universit\'e Paul Sabatier,
  F-31062 Toulouse, France
}

\date{June 16, 2004}
\title{Spin-Charge Separation in Two-dimensional Frustrated Quantum Magnets}
\begin{abstract}
  The dynamics of a mobile hole in two-dimensional frustrated
  quantum magnets is investigated by exact diagonalization techniques. Our results
  provide evidence for spin-charge separation upon doping the
  \kag lattice, a prototype of a spin liquid. In contrast, in
  the checkerboard lattice, a symmetry broken Valence Bond Crystal, 
  a small quasi-particle peak is seen for some crystal
  momenta, a finding interpreted as a restoration of weak
  holon-spinon confinement.
\end{abstract}
\maketitle

\begin{figure}
  \centerline{\includegraphics*[width=0.9\linewidth]{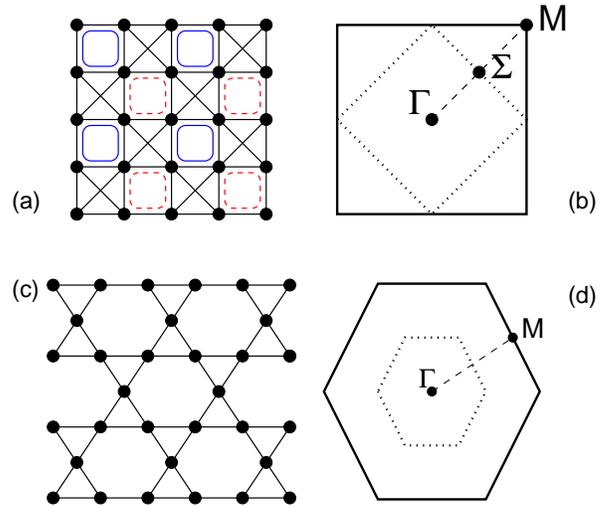}}
  \caption{\label{fig:Lattices}
    (a) Schematic picture of the VBC GS of the Heisenberg model on the
    checkerboard lattice. The two degenerate singlet patterns are shown 
    with full blue and dashed red lines. 
    (b) Dotted line: Brillouin zone (BZ) of the checkerboard lattice.
    Full line: extended BZ equivalent to the square lattice BZ.
    The path from the zone center $\Gamma$ to the ${\bf k}=(\pi,\pi)$
    point $M$ is shown as a dashed line. 
    (c) The \kag lattice; 
    (d) Dotted line: BZ of the \kag lattice.
    Full line: extended BZ equivalent to the nondepleted triangular lattice BZ.
  }
\end{figure}

In the last years, there has been growing interest in the
investigation of frustrated magnets, both on experimental and
theoretical sides. Spin-1/2 Heisenberg models on two and three
dimensional (3D) lattices have been studied extensively providing
increasing evidences for new exotic phases like the spin
liquid Resonating Valence Bond (RVB)~\cite{RVB} state or the
Valence Bond Crystal (VBC)~\cite{VBC}. In two dimension (2D), the
checkerboard lattice~\cite{pyrochlore2D,Fouet,Brenig}, i.e.~the 2D analog of the
3D pyrochlore lattice, and the \kag lattice~\cite{KagomeED,KagomeED_2}
(see Fig.~\ref{fig:Lattices}), respectively, seem to provide simple 
theoretical realizations of VBC and RVB groundstates (GS). 
In addition, an exotic scenario where magnons break up into ($S=1/2$)
deconfined spinons~\cite{spinons} could take place in a RVB state. These
theoretical conjectures have triggered a renewal of experimental
activity on quantum frustrated magnets like e.g.~the ($S=3/2$)
\kag compound SrCr$_{9p}$Ga$_{12-9p}$O$_{19}$~\cite{SCGO} and
the pyrochlore titanates RE$_2$Ti$_2$O$_7$ (where RE is a
rare-earth magnetic atom such as gadolinium or
terbium)~\cite{titanates}. 
These materials exhibit very rich phase diagrams as a function of 
chemical composition, temperature, pressure and magnetic field which
underline the crucial role of geo\-metric frustration.
Furthermore, the recent discovery of superconductivity in a CoO$_2$ based
compound with triangular layers~\cite{cobaltites} has revived interest in 
exotic RVB mechanisms of superconductivity~\cite{Baskaran}.

In this Letter, we investigate the dynamics of a mobile hole
injected in various 2D frustrated quantum magnets. This study is
indirectly motivated by the long-standing puzzles offered by the
``normal phase'' of high-$T_c$ superconductors whose discovery
initiated a search for spin-charge separation mechanisms in 2D and
the first proposal of 2D Luttinger liquids~\cite{2D-LL}. The
dynamics of a mobile hole doped into an (Ising) antiferromagnet
(AF) was first studied by Brinkman and Rice~\cite{BrinkmanRice}
and a coherent motion was later predicted by Trugman~\cite{Trugman}.
However, a close inspection of the hole spectral function showed
evidences for a composite nature of the doped hole quasiparticle
(QP) showing holon and spinon components~\cite{Laughlin}. Large
quantum fluctuations present in frustrated magnets might then
induce a complete deconfinement of the hole components. Here we
present large scale exact diagonalization results for single
hole spectral functions and hole-spin correlations obtained on the
checkerboard and the \kag antiferromagnets. A finite lifetime
of the hole (breaking apart into a spin 1/2 spinon and a charge
$Q=e$ holon) is signaled by the absence of a QP peak in the
computed spectral functions. Since the latter could, in principle,
be measured by Angular Resolved Photoemission Spectroscopy (ARPES)
on high-quality (cleaved) {\it insulating} single
crystals~\cite{SrCuOCl}, our results can provide motivation and
theoretical background to experimentalists. In addition, this
investigation might also be relevant to lightly doped samples.
Evidences for spin-charge separation in the \kag spin liquid GS are
provided while, in the checkerboard lattice, we find small QP peaks
for some momenta.

Before going further, let us first summarize briefly the
properties of the undoped insulating systems we are
dealing with. The checkerboard lattice is believed to form
a VBC with plaquette singlets on a subset of the
void plaquettes \cite{pyrochlore2D,Fouet,AltmanAuerbach,note_j1j2}. 
The groundstate manifold
is two fold degenerate [see Fig.~\ref{fig:Lattices} (a)]. The system
has quite a sizable triplet gap $\Delta_T\approx 0.7 J$~\cite{Fouet,Brenig},
and a somewhat smaller singlet gap $\Delta_S\approx 0.3 J$~\cite{Fouet,AltmanAuerbach}
to domain wall like singlet excitations~\cite{AltmanAuerbach}. Based on
this picture of a fully gapped system, we expect a coherent
QP motion of an injected hole at least in parts of the Brillouin zone (BZ)
as observed for example in Heisenberg ladders or dimerized spin chains.
On the other hand, the undoped \kag lattice has a very interesting and puzzling
groundstate of a new type: while this system is also magnetically
disordered \cite{KagomeED}, it has quite a small triplet gap
($\Delta_T \approx 0.05 J$)~\cite{KagomeED_2}. The singlet 
sector seems to remain gapless showing an exponentially large 
number of singlets within the spin gap~\cite{KagomeED_2,KagomeSinglets}.
These unconventional low lying excitations open the door to new and
surprising phenomena upon hole doping.

Our numerical calculations are based on the standard $t{-}J$ model Hamiltonian:
\begin{equation}
  - t \sum_{\langle i,j\rangle,\sigma}\ \mathcal{P} \left(
    c^{\dagger}_{i,\sigma}c_{j,\sigma} +\mbox{h.c.}\right) \mathcal{P}
  + J \sum_{\langle i,j\rangle}\ \S_i \cdot \S_j -\frac{1}{4} n_i n_j
\end{equation}
where on both lattices all bonds have the same couplings $t$ and $J$.
This model is believed to be reliable to describe weakly doped Mott-Hubbard 
insulators with large optical gaps.
Hereafter, a physical value of $J/|t|=0.4$ is assumed, unless mentioned otherwise.
The spectral functions are defined in the standard way:
\begin{equation}
  A^{\sigma}({\bf k},\omega)=-\frac{1}{\pi}
  \mbox{Im}[\langle\Psi_0|c^\dagger_{{\bf k},\sigma}
  \frac{1}{\omega+E_0+i\eta-H}c_{{\bf k},\sigma}|\Psi_0\rangle],
\end{equation}
and calculated by the Lanczos continued-fraction technique on finite clusters
of up to 32 sites with periodic boundary conditions. Here $|\Psi_0\rangle$ 
(of energy $E_0$) is the GS of the undoped insulating system, so that the 
dynamics of a {\it single} hole is probed. A simple sum rule is satisfied:
$\int A^{\sigma}({\bf k},\omega) \mbox{d}\omega=N_{\sigma}/N 
\simeq 1/2$, independent of ${\bf k}$. Therefore the spectral function plots 
share an arbitrary, but common scale, unless mentioned otherwise.
In addition an artificial broadening $\eta=0.05|t| \mbox{ or } 0.1|t| $
has been used to plot the data. The weights of the lowest peaks can be obtained 
exactly, independent of $\eta$. We have indicated this weight in the plots when 
a QP peak is present.

\begin{figure}
  \centerline{\includegraphics*[width=0.98\linewidth]{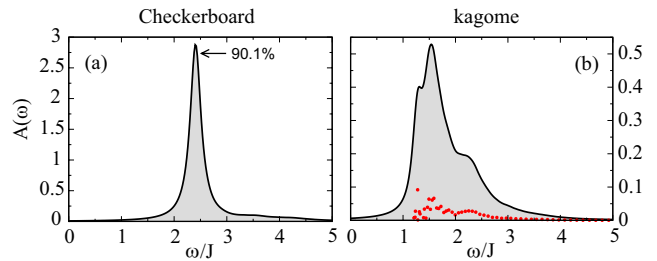}}
  \caption{\label{fig:spectral_static}
    Single hole spectral function for a static ($t=0$) hole in a checkerboard
    (a) or a \kag (b) lattice. 
    The checkerboard spectral function is dominated by the lowest 
    energy peak, in contrast to the \kag response. For this case
    the weight of the poles is indicated by red circles.
}
\end{figure}

As a first step it is of special interest to consider the static limit i.e.~the
case $t=0$ (equivalent to an infinite hole mass) where a spin
1/2 is replaced by an inert (impurity-like) site. The spectral
function, which becomes here independent of momentum ${\bf k}$,
gives useful hints on the host magnetic disturbance induced by the
static hole. Results in Fig.~\ref{fig:spectral_static} show very different
qualitative behaviors for the checkerboard and the \kag lattices.
A sharp peak which exhausts most of the spectral weight is found in
the checkerboard lattice. This is consistent with the picture of
breaking a local plaquette singlet, therefore producing a localized
moment, which remains confined to the hole. The very large peak weight
shows that the bare state $c_{i,\sigma}|\Psi_0\rangle$ is a
very good approximation of the true one hole groundstate.
This however seems to be quite different for the \kag case: the spectral 
function shows a rather broad response, with no pronounced low energy peak [see weight
distribution (circles) in Fig.~\ref{fig:spectral_static} (b)]. 
Here the overlap of the bare electron removal state and the true groundstate of the one
impurity problem is tiny. This can be nicely understood in the
``dimer freezing'' picture advocated in Ref.~\cite{StaticKagomeImpurities},
where the screening of the impurity constitutes an important deformation
of the RVB spin liquid groundstate. This suggests that a single dopant in
the \kag lattice generates a dimer-screened holon with charge $Q$=$e$ and spin $S$=0
and a deconfined spinon ($Q$=0, $S$=1/2).

\begin{figure}
  \centerline{\includegraphics*[width=0.9\linewidth]{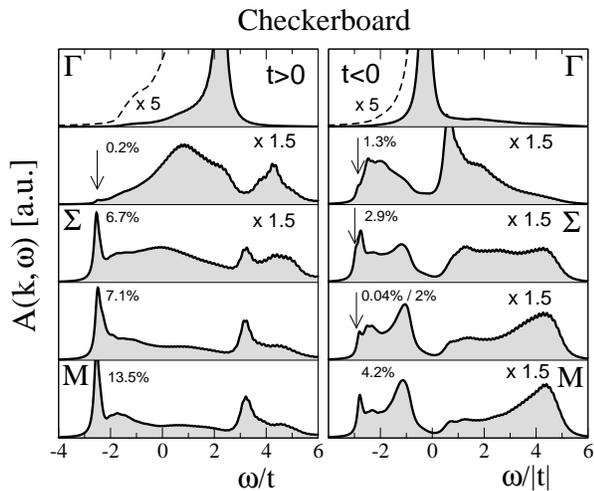}}
  \caption{\label{fig:SpectralCheckerboard}
    Single hole spectral functions obtained on a 32 site checkerboard cluster
    along the line $\Gamma \leftrightarrow M$.
    Left panel $t=+1$, right panel $t=-1$. In both cases $J/|t|=0.4$.
    When a quasiparticle peak is present the corresponding weight is indicated.
    Scaling factors are applied as indicated on the plots. \vspace{-4mm}
}
\end{figure}

We now move to the case of a dynamic hole ($t\ne 0$).
Because of the absence of particle-hole symmetry in frustrated
lattices one has to distinguish between $t>0$ and $t<0$. Note that
for $t<0$ frustration can also appear in the hole motion. For
example, a tight-binding particle on an isolated triangle gains a
kinetic energy $|t|$ per particle, a factor of two smaller than
for $t>0$. Our results for a 32 cluster \cite{clustershape} are shown
in Fig.~\ref{fig:SpectralCheckerboard} along the $\Gamma \leftrightarrow
M$ line of the BZ [Fig.\ref{fig:Lattices} (b)]. In all cases most 
of the spectral weight is found to be incoherent, distributed over a 
range of $7-9\ |t|$. However, a small QP peak is visible, in particular
for momenta close to the $M$-point.
The region close to the $\Gamma$ point has a tiny or no QP peak and
the shape of the spectral function at $\Gamma$ itself is very special, 
probably because of a higher point group symmetry. It is interesting to
notice, in the case $t<0$, the presence of a pseudo-gap in the
region $\omega\sim 0$. 

\begin{figure}
  \centerline{\includegraphics*[width=0.9\linewidth]{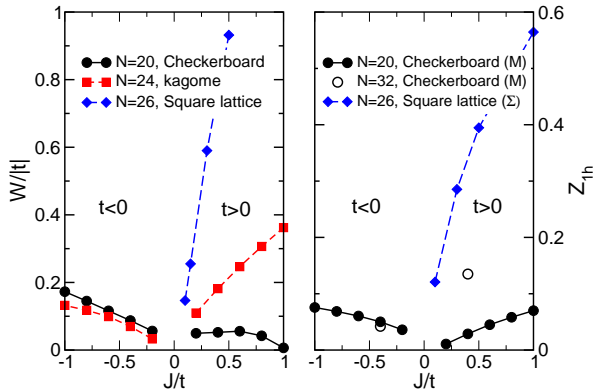}}
  \caption{\label{fig:Data_vsJ}
    Left panel: QP bandwidth in the checkerboard lattice vs $J/t$ $(J>0)$.
    Comparison is shown with the case of the square lattice (data from 
    Ref.~\protect\cite{Laughlin}).
    The magnitude of the dispersion of the low-energy edge of the \kag one-hole
    spectrum is also shown. 
    Right panel: checkerboard QP weight at the $M$ point of the BZ vs $J/|t|$
    and square lattice QP weight at the $\Sigma=(\pi/2,\pi/2)$ point.
    \vspace{-4mm}
  }
\end{figure}

We have estimated the QP bandwidth $W$ of the checkerboard lattice
from the dispersion of its QP pole. As shown in the left panel
of Fig.~\ref{fig:Data_vsJ} the single hole bandwidth is much
smaller than in a 2-leg spin ladder, which has a gapped spectrum, 
($W\sim 2t$ for strong magnetic rung couplings~\cite{ladder}) 
or even in the square lattice, which has a gapless spectrum,
where $W\sim 2.2J$. In the latter case,
the strong renormalization of the coherent motion can be explained
e.g.~by long-wavelength spin-waves scattering~\cite{Trugman,Laughlin}.
In the checkerboard lattice the QP are more massive, almost localized,
although the VBC host has no low energy excitations. We believe that this
remarkable fact is due to a destructive interference effect between the paths
available for the hole to hop from one plaquette to the next. One
can define QP weights $Z_{\mbox{\tiny 1h}}({\bf k})$ as the relative weights
contained in the first $\delta$-function appearing at the bottom
of each spectrum in Fig.~\ref{fig:SpectralCheckerboard}. Its
behavior versus $J/t$ is shown in the right panel of Fig.~\ref{fig:Data_vsJ}
and compared with existing data for the square lattice. We note that,
although there are quantitative differences, the qualitative behavior 
is similar; larger $J$ leads to larger holon-spinon confinement and 
hence to an increase of $Z_{\mbox{\tiny 1h}}$.

\begin{figure}
  \centerline{\includegraphics*[width=0.9\linewidth]{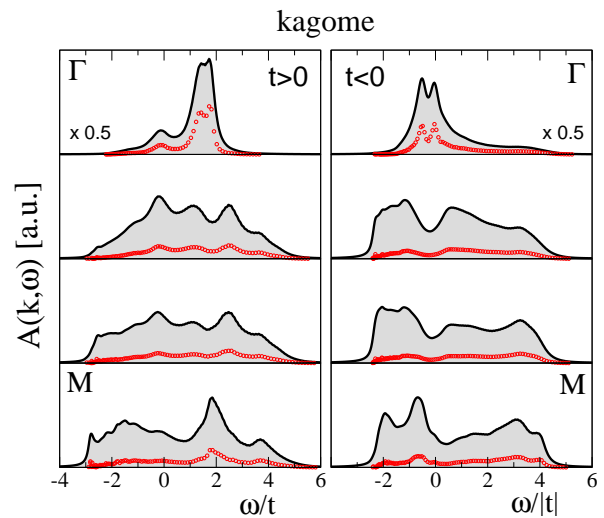}}
  \caption{\label{fig:Kagome27}
    Single hole spectral functions (black lines) along the
    line $\Gamma \leftrightarrow M$ computed on a 27 site \kag 
    cluster for $t=+1$ (left panel) and $t=-1$ (right panel).
    In both cases $J/|t|=0.4$. The red circles denote pole locations
    and their residues. Note that no quasiparticle peaks are visible for 
    all momenta.
    \vspace{-4mm}
  }
\end{figure}

The spectral functions of the \kag lattice shown in
Fig.~\ref{fig:Kagome27} show exotic behaviors; they are very broad
for all momenta (width $\sim 6-8\ |t|$) and, in contrast to the 
checkerboard lattice, {\it do not show visible QP peaks} both for 
$t>0$ (left panel) and $t<0$ (right panel). Note that the broad
appearance is not due to a large $\eta$, but is an intrinsic feature
of the spectral function as can be seen from the large number of poles
carrying spectral weight (circles in Fig.~\ref{fig:Kagome27}).
Note that the GS $|\Psi_0\rangle$ of the 27 site undoped \kag cluster
is a spin  1/2 state, e.g.~$S^Z=+1/2$, $\mid\uparrow\rangle$. Therefore, the initial 
state of charge $Q=e$ and spin projection $S^Z=0$, 
$c_{{\bf k},\uparrow}\mid\uparrow\rangle$, is an equal-weight superposition of
a singlet and a triplet state leading to two contributions.
We have checked that both components are consistent with the decay 
of the original hole into a holon and a spinon (with no QP peak). 
Close similarities of the singlet and triplet channels also suggest that 
the spinons in the final state (with either parallel or anti-parallel spins)
are weakly interacting in agreement with the mechanism of spinon deconfinement.

\begin{figure}
  \centerline{\includegraphics[width=0.9\linewidth]{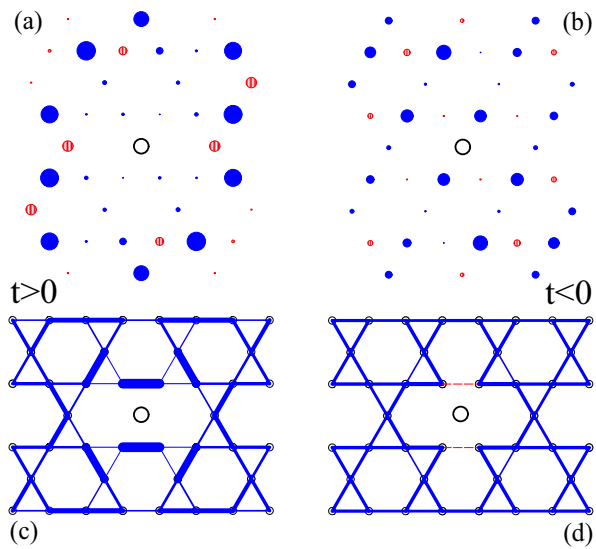}}
  \caption{\label{fig:Correlations}
    (a-b) Hole-spin density-density correlation function 
    $\langle n_i^h S_j^Z\rangle$ calculated on
    a 30 site \kag cluster for both signs of $t$. The diameter
    is proportional to the magnitude of the correlation. Full blue
    (striped red) circles denote positive (negative) correlations.
    (c-d) Nearest neighbor bond spin-spin correlations 
    around the mobile hole computed on a 27 site sample 
    for both signs of $t$. $J=0.4$ for all the plots.
    Full blue lines denote antiferromagnetic
    correlations, while the red, dashed lines stand
    for ferromagnetic correlations.
    \vspace{-4mm}
  }
\end{figure}

The spectral function data strongly support a spin-charge
separation scenario for the \kag lattice. In order to get
further evidence and more insight about the origin of this
phenomenon we also investigate the spin density profile in
the vicinity of the hole, i.e.~the hole-spin correlation function
$\langle n_i^h S_j^Z\rangle$ in the one hole-doped 30 site
\kag cluster (for which $S_Z=1/2$). Results are shown in
Fig.~\ref{fig:Correlations} for $t=+1$ (a) and $t=-1$ (b).
In the case $t=+1$ we clearly see a repulsion between the net
$S$=1/2 moment and the mobile hole. The case $t=-1$ displays a 
different picture, showing a rather uncorrelated behavior
between the spinon and the holon. Nevertheless a spinon-holon
confinement does not seem to be present, in agreement with
spectral function data. In contrast we have checked that 
on a checkerboard lattice a clear localization of the spinon 
close to the holon takes place.

Finally we consider the GS of the one hole-doped 27 site
\kag cluster of total charge $Q=e$ and spin $S=0$ to mimic the
holon wavefunction. Figs.~\ref{fig:Correlations} (c,d) show a
snapshot of the nearest neighbor spin correlations around
the {\em mobile} holon. Surprisingly in the case $t=+1$ a 
``dimer freezing'' is observed very similar to the {\em static} 
impurities considered in Ref.~\cite{StaticKagomeImpurities}. In 
contrast, the case $t=-1$ shows two slightly ferromagnetic bonds
on the two triangles containing the holon, while the other bonds 
remain largely unaffected. This behavior of the magnetic environment 
is fairly well described by a simple energetic consideration on a 
3-site ring $t{-}J$ model; for $t=+1$ the propagating holon is 
accompanied by a singlet, while in the opposite $t=-1$ case an 
accompanying ferromagnetic bond appears.

To conclude, based on extensive analysis of the single particle spectral
functions and various charge-spin correlations in the single hole GS, we have
addressed the issue of spin-charge separation in 2D magnetically
disordered magnets. We show strong evidence that it occurs in the \kag 
lattice and we expect it to be robust for small but finite doping.
This provides the first example of observed spin-charge separation in
a two-dimensional microscopic model. The RVB nature of the undoped 
groundstate (characterized by a gapless low-excitation singlet spectrum)
seems to be crucial for this behavior. Indeed, in the checkerboard lattice, 
which exhibits a VBC structure, a weak hole-spinon confinement manifests itself
as QP peaks for some momenta.

\acknowledgments
We thank M.~Mambrini for useful discussions. A.L.~acknowledges
support from the Swiss National Fund and the CNRS. D.P.~thanks the
Institute for Theoretical Physics, ETH Z\"urich, for hospitality during
the final stage of this work. We thank IDRIS (Orsay) and the LRZ M\"unchen 
for allocation of CPU-time.
 

\begin{thebibliography}{}
\bibitem{clustershape}
  We use the cluster $32^*$ as defined in Ref.~\cite{Fouet}.
\end{thebibliography}
\end{document}